\documentclass[conference,final]{IEEEtran}

\usepackage{booktabs} %
\usepackage{listings}
\usepackage{fancyvrb}
\usepackage{wrapfig}
\usepackage{graphicx}
\usepackage{amsmath}
\usepackage{amssymb}
\usepackage{color}
\usepackage{ifpdf}
\usepackage{subcaption}
\usepackage{float}
\usepackage[utf8]{inputenc}
\usepackage{multirow}
\usepackage{rotating}

\usepackage{xcolor}
\PassOptionsToPackage{hyphens}{url}
\usepackage[colorlinks,breaklinks=true]{hyperref}
\AtBeginDocument{%
  \hypersetup{
    citecolor=blue,
    linkcolor=blue,   
    urlcolor=blue}}

\usepackage{url}
\usepackage{booktabs}
\usepackage{listings}
\usepackage{paralist} 
\usepackage{wrapfig}
\usepackage{orcidlink}

\usepackage{multirow}
\usepackage{ifpdf}
\usepackage{xspace}
\usepackage{keyval}
\usepackage{color}
\usepackage[font=small,labelfont=bf]{caption}

\definecolor{listinggray}{gray}{0.95}
\definecolor{darkgray}{gray}{0.7}
\definecolor{commentgreen}{rgb}{0, 0.4, 0}
\definecolor{darkblue}{rgb}{0, 0, 0.4}
\definecolor{middleblue}{rgb}{0, 0, 0.7}
\definecolor{darkred}{rgb}{0.4, 0, 0}
\definecolor{brown}{rgb}{0.5, 0.5, 0}

\usepackage[normalem]{ulem}
\makeatletter
\def\cyanuwave{\bgroup \markoverwith{\lower3.5\p@\hbox{\sixly \textcolor{cyan}{\char58}}}\ULon}
\def\reduwave{\bgroup \markoverwith{\lower3.5\p@\hbox{\sixly \textcolor{red}{\char58}}}\ULon}
\def\blueuwave{\bgroup \markoverwith{\lower3.5\p@\hbox{\sixly \textcolor{blue}{\char58}}}\ULon}
\font\sixly=lasy6 %
\makeatother

\newif\ifconceptualmodel
\conceptualmodelfalse

\newif\ifkeepeverything
\keepeverythingfalse

\newif\ifthroughputemulator
\throughputemulatorfalse

\newcommand{\orcid}[1]{\href{https://orcid.org/#1}{\textcolor[HTML]{A6CE39}{\aiOrcid}}}

\newcommand{\pilotabstraction}{pilot abstraction\xspace}

\newcommand{\radicaldreamer}{RADICAL-DREAMER\xspace}

\newcommand{\up}{\vspace*{-1em}}
\newcommand{\upp}{\vspace*{-0.5em}}

\lstdefinestyle{myListing}{
  frame=single,
  backgroundcolor=\color{listinggray},
  language=C,
  basicstyle=\ttfamily \footnotesize,
  breakautoindent=true,
  breaklines=true
  tabsize=2,
  captionpos=b,
  aboveskip=0em,
  belowskip=-2em,
}

\lstdefinestyle{myPythonListing}{
  frame=single,
  backgroundcolor=\color{listinggray},
  language=Python,
  basicstyle=\ttfamily \scriptsize,
  breakautoindent=true,
  breaklines=true
  tabsize=2,
  captionpos=b,
}

\ifpdf
\DeclareGraphicsExtensions{.pdf, .jpg, .tif}
\else
\DeclareGraphicsExtensions{.ps,  .eps, .jpg}
\fi

\usepackage{listings}
\usepackage{paralist}

\title{Exploring Task Placement for Edge-to-Cloud Applications using Emulation\upp\upp} 
\author{\\Andre Luckow$^{1,2,3, \orcidlink{0000-0002-1225-4062}}$, Kartik Rattan$^{1}$, Shantenu Jha$^{4,1}$\\
   \footnotesize{\emph{$^{1}$RADICAL, ECE, Rutgers University, Piscataway, NJ 08854, USA}}\\
   \footnotesize{\emph{$^{2}$Ludwig-Maximilian University, Munich, Germany}}\\
   \footnotesize{\emph{$^{3}$Clemson University, South Carolina, USA}}\\
   \footnotesize{\emph{$^{4}$Brookhaven National Laboratory, Upton, NY, USA}\upp\upp\upp\up}
}

\begin{document}

\date{}
\IEEEoverridecommandlockouts
\IEEEpubid{\makebox[\columnwidth]{ xxx~\copyright2020 IEEE \hfill} \hspace{\columnsep}\makebox[\columnwidth]{ }}
\maketitle
\IEEEpubidadjcol

\begin{abstract}

A vast and growing number of IoT applications connect physical
devices, such as scientific instruments, technical equipment, machines, and
cameras, across heterogenous infrastructure from the edge to the cloud to
provide responsive, intelligent services while complying with privacy and security
requirements.
However, the integration of heterogeneous IoT, edge, and cloud technologies
and the design of end-to-end applications that seamlessly work across multiple
layers and types of infrastructures is challenging. A significant issue is
resource management and the need to ensure that the right type and scale of
resources is allocated on every layer to fulfill the application's processing
needs. As edge and cloud layers are increasingly tightly integrated,
imbalanced resource allocations and sub-optimally placed tasks can quickly
deteriorate the overall system performance. This paper proposes an emulation
approach for the investigation of task placements across the edge-to-cloud
continuum. We demonstrate that emulation can address the complexity and many
degrees-of-freedom of the problem, allowing us to investigate essential
deployment patterns and trade-offs.  We evaluate our approach using a machine
learning-based workload, demonstrating the validity by comparing emulation and
real-world experiments. Further, we show that the right task placement
strategy has a significant impact on performance -- in our experiments,
between 5\,\% and 65\,\% depending on the scenario.

\end{abstract}


\section{Introduction}\label{sec1}

The \emph{Internet-of-Things (IoT)} is driving the current data deluge to
unprecedented scales. Howell~\cite{iot_2030} predicts that the number of IoT
devices worldwide will grow annually by 12\,\% to 125 billion in 2030. As
these devices continuously produce data, the need to process and respond to
this data in real-time is increasing. IoT devices and real-world processes often need to tightly integrate to enhance situational awareness, enable real-time decisions, and the steering and control of devices.  Thus, a
timely response is often essential. IoT has become the driving force behind
the development of distributed infrastructures for data generated by
scientific experiments and observatories, such as light sources~\cite{nsls},
earth observatories~\cite{Geernaert},  astronomy
observatories~\cite{Berriman:2011:AAS:2039359.2047483}, and the Large Hadron
Collider~\cite{lhc}.

While centralized cloud and high-performance computing (HPC) infrastructures
address many computational requirements, they have limitations~\cite{7543455}.
In particular, IoT scenarios require the management of voluminous data and are
challenged by latencies, bandwidths, privacy, sovereignty, and cost
requirements. Simultaneously, the types of infrastructures in the continuum 
are diversifying, e.\,g.,  increasingly powerful and heterogeneous 
IoT devices~\cite{futureMLtiny, MovingToEdge},
gateway servers, local compute infrastructure, and centralized clouds. This
landscape of infrastructure is commonly referred to as the \emph{edge-to-cloud
continuum  (abbrev. continuum)}. It opens the possibility of bringing processing
closer to the data source, thereby addressing the mentioned
challenges~\cite{harnessing}.

However, utilizing and exploiting the edge-to-cloud continuum capabilities is
hindered by a lack of understanding of application requirements and patterns,
particularly concerning infrastructures and resource management. This paper
provides an in-depth characterization of edge-to-cloud applications, studying
common application characteristics and deployment modalities. Although
edge-to-cloud applications often vary in many technical details, they share
commonalities, e.\,g., in how they deploy and distribute machine learning
pipelines across the continuum. Our analysis supports both application and tools
developers in designing and building applications and tools that build on these characteristics and
patterns.

Distributed resource and task management are crucial to satisfy applications'
performance and scalability requirements.  There are important questions that
applications need to consider for optimizing their task
placements: How to partition an application workload across multiple layers of
infrastructures? How many resources should be allocated on what layer? How
should the application respond to changes in the available resources? 

Thus, the ability to explore important task management trade-offs along the
edge-to-cloud continuum is crucial. Often, only a subset of these factors can be
evaluated using experiments. Extensive experimentation is often
prevented by the availability of infrastructure and not time- and
cost-effective. An important method that addresses these challenges is
emulation, which allows the investigation of different aspects of IoT systems
with a significantly reduced effort.

We utilize the \radicaldreamer~\cite{radical_dreamer} emulation framework for
modeling task placement in edge-to-cloud applications. Particularly, we evaluate
workload and resource management aspects of edge-to-cloud applications, e.\,g.,
effective strategies for distributing workloads. Using the framework, we
assess important metrics, e.\,g., time-to-completions broken down by different
processing stages. We demonstrate our approach using extensive emulation
experiments on XSEDE to investigate workload and task placements 
for different deployment modalities.

\begin{table*}[th]
  \footnotesize
	\centering
	\begin{tabular}{|p{2.3cm}|p{4.5cm}|p{4.5cm}|p{5.2cm}|}\hline
	   				    &\textbf{Farming} &\textbf{Manufacturing}		&\textbf{Scientific Experiments}\\\hline   
\textbf{Description} &Collection of spatial, temporal sensor data from the farm to enable optimizations, e.\,g., water, fertilizer usage. 
			& Usage of IoT and machine data for improving process efficiencies, e.\,g., predictive maintenance and supply chain.
			&Usage of internet-connected instruments in different science domains, e.\,g., astronomy, high-energy physics, light source sciences \\\hline
		
\textbf{Sensing}     &Electric conductivity, temperature,	
	   		microphones, cameras, spectrometers, weather    &Logs, errors, vibration, temperature       &Highly-specialized sensors built for specific experiments, high data rates, e.\,g., light source (up to 20\,GB/sec)\\\hline 

\textbf{Data} 	&Time series sensor data, sound, images &Time Series sensor data, sound, image &highly specialized data, e.\,g., light source raw data, astronomy. High volumes of historical data. High data rates during campaigns. \\\hline 

\textbf{Processing}   &Outlier detection, feature extraction, image recognition, sound detection, anomaly 
			detection           
			&Outlier detection, image classification, object detection
			&Outlier detection, image reconstruction, image classification, object detection\\\hline 

\textbf{Edge-to-Cloud Continuum} &Edge: sensing \& data collection, compression, monitoring \linebreak[4]
                         Cloud: analytics, machine learning
                        &Edge: sensing \& data collection, compression, monitoring, process control  \linebreak[4]
                        Cloud: analytic, machine learning   
                        &Edge: sensing \& data collection, compression,monitoring, steering\linebreak[4]
                        Cloud: analytics, machine learning, simulation, optimization  \\\hline   

\textbf{Data Exchange} &Edge-to-Cloud: Send only relevant data from edge-to-cloud      
                       &Edge-to-Cloud: Send meta-data and selected raw data for ML training selected amount from edge to cloud.  \linebreak[4]
                       Cloud-to-Edge: ML inference results
                       &Edge-to-Cloud: Send meta-data and selected raw data for ML training selected amount from edge to cloud.  \linebreak[4]
                       Cloud-to-Edge: ML inference and simulation results for steering and control \\ \hline

	   \hline
	\end{tabular} 
	\caption{\textbf{Edge-to-Cloud Application Domains:} While IoT application domains are highly diverse, applications share the need for processing sensor data (e.\,g., image, sound)  across multiple tiers of computing and data infrastructure.\label{tab:application_scenarios}\up}
\end{table*}

This paper is structured as follows: We discuss application scenarios and
characteristics in section~\ref{sec:applications}.
Section~\ref{sec:workload_emulator} describes the emulation approach. We provide
an extensive evaluation in section~\ref{sec:experiments}. Finally,
section~\ref{sec:related} discusses related work.

\section{Applications Scenarios and Requirements}
\label{sec:applications}

Edge-to-cloud applications are highly diverse. Data scales, rates, and usage
modes (edge-only, cloud-only, hybrid/edge-to-cloud) can vary significantly.
However, they share many commonalities, e.\,g., the need to use the continuum to
address performance, security, and privacy.  We focus on IoT applications,
i.\,e., applications that manage complex data processing pipelines across the
edge-to-cloud continuum and discuss common patterns. We analyze different
application scenarios from different application domains with the objective to
identify common patterns.

\subsection{Applications Scenarios}
\label{sec:app_scenarios}

IoT devices are increasingly deployed in many scientific and industrial domains
to allow data collection and analytics, improve process efficiencies, support
remote operations and real-time control. Examples are large-scale scientific
instruments and observatories found in many sciences (e.\,g., astrophysics,
earth sciences), health care, farming, energy, mobility, factories, and supply
chain (see Chabas et al.~\cite{mck_edge}). In the following, we investigate
three application domains: farming, manufacturing, and scientific experiments.

\emph{Agriculture:} IoT and data analytics help understand essential aspects
of agriculture, e.\,g., energy,  water usage, and yields~\cite{SmartFarm17}.
An example is the usage of weather data and IoT data from the vineyard, such
as environmental conditions and wine stress, to better guide  the water
irrigation system of the wine yard, reducing water usage~\cite{iot_wine}. Data
collected from farms can be highly heterogeneous, e.\,g., spectral data for
monitoring wine fermentation process, sound data for localization of
livestock, and image data from cameras, drones, satellites.

\emph{Manufacturing:} IoT sensors are also increasingly used in manufacturing
environments. Data collected in such environments is crucial for use cases,
such as predictive maintenance and visual inspection, planning, and
optimization~\cite{8622357,doi:10.1002/spe.2816}.

\emph{Scientific Experiments and Facilities:} Edge and fog capabilities are
increasingly important for many scientific experiments, e.\,g., synchrotron
light source experiments. Experimental facilities like the National Synchrotron Light Source II (NSLS-II) or the X-Ray Free Electron Laser (XFEL) are  generating data at growing rates of up to 20\,GB/sec~\cite{nsls}. This data often needs to be processed in a
time-sensitive fashion, e.\,g., to either steer the experiments or update its digital twin~\cite{lcls_data}. In earth
and environmental sciences, edge computing is a crucial enabler for increasing
the impact of observatories. For example, the Atmospheric Radiation
Measurement (ARM) facility  envisions the in situ coupling of ARM observations
with atmospheric and climate models on the edge to accelerate the
time-to-insight~\cite{Geernaert}.

\subsection{Application Characteristics}

In Table~\ref{tab:application_scenarios}, we use five categories to characterize application scenarios: sensing, data, processing, edge-to-cloud usage,  and data exchange.

\emph{Sensing} describes the process of capturing changes in the environment using sensors, e.\,g., for images, sound, temperature. Often, a specialized embedded CPU is used to perform this task. Often, some basic pre-processing is conducted during the sensing process to extract relevant events from the time-series raw sensor data. The \emph{data} category describes the data characteristics, in particular, data types.

\emph{Processing} characterizes the algorithms used for extracting insight from data, e.\,g., machine learning and analytics algorithms. Depending on the used approach, e.\,g., analytics with simple statistics or deep learning, different computational characteristics arise~\cite{NAP18374,bigdata-ogres}. In the context of machine learning, the processing characteristic can be differentiated in model training and inference. Further,  \emph{pre-processing} is essential to reduce the data volume and prepare for further processing, e.\,g., pre-aggregation, normalization, and sampling. 

Table~\ref{tab:characteristics_stages} describes the characteristics of the different types of processing workloads: pre-processing, analytics, inference and training. Depending on the choice of algorithm, the computational demands can vary significantly. A well-balanced system requires careful tuning of compute resources, bandwidths and algorithms. 

\begin{table}[t]
	\centering
	\begin{tabular}{|p{2.2cm}|p{1cm}|p{1cm}|p{1cm}|}\hline
								&\textbf{Compute} &\textbf{Memory} &\textbf{I/O} \\\hline
		\textbf{Pre-processing} 			&o       &+		     &++ \\\hline
		\textbf{Analytics} 				&+  	 &+    	     &+  \\\hline
		\textbf{Inference} 				&++  	 &++    	 &+  \\\hline
		\textbf{Training} 				&+++ 	 &+++        &+  \\\hline 
	\end{tabular}
	\caption{Typical Characteristics of the Different Types of Processing Tasks \label{tab:characteristics_stages} }
\end{table}

\emph{Edge-to-Cloud} describes how the applications is utilizing cloud and edge resources.  Cloud capabilities allow for more intelligence (global analytics), improved model quality through the ability to use more complex models, and performance (more data and compute available). \emph{Data exchange} characterizes data transmission patterns along the continuum.

\subsection{Towards Application Patterns}
\label{sec:app_discussion}

\begin{figure}[t]	
  \centering
	\includegraphics[width=0.5\textwidth]{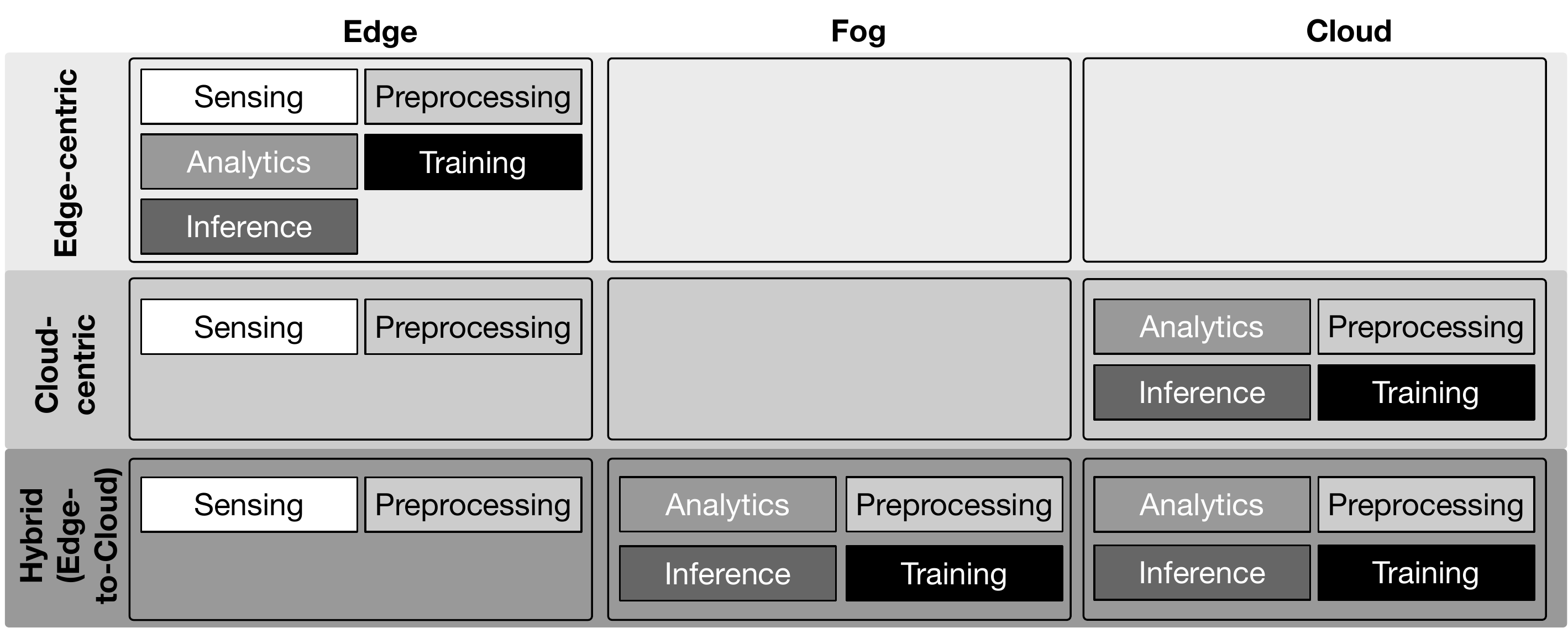}
	\caption{\textbf{Deployment Modalities for Edge-to-Cloud Applications:} While traditional deployment scenarios are either cloud or edge-centric, increasingly, it is required to use more complex edge-to-cloud architectures to address performance, security, and privacy.  \label{fig:edge-to-cloud-deployment} \up\up}
\end{figure}

While the application scenarios  differ significantly, they share
commonalities concerning the need to process sensor data, e.\,g., image, and
sound, across multiple tiers of infrastructures. A re-occurring pattern is processing this data across multiple hierarchical layers of edge and cloud resources~\cite{simmhan:iotn:2017}.

Figure~\ref{fig:edge-to-cloud-deployment} illustrates three common deployment
modalities: cloud-, edge-centric and hybrid for distributing processing tasks
across the continuum. In the cloud-centric modality, most services and tasks are run in the cloud; in the edge-centric modality on the edge.  
However, depending on the application scenario and available
infrastructure, it is often beneficial to distribute tasks across
the continuum utilizing a hybrid deployment modality. 

On the edge, the focus is often on sensing, pre-processing, and compression; the
cloud and intermediate layers (also referred to as fog) compute more advanced,
global data aggregations, including machine learning inference and modeling.
For example, while real-time applications often require inference on the edge to
ensure low latencies, these edge capabilities often benefit from up-to-date
models  trained and updated using data collected from multiple devices
using advanced computational resources in the cloud. Another hybrid usage mode
is the cloud usage for processing regions of interest identified using an edge-capable model, e.\,g., an anomaly detector. Often, cloud processing can accommodate  
more compute-intensive high-fidelity models.

Defining a suitable task placement strategy  requires the consideration of various
application and infrastructure characteristics and their interactions. In the following, we explore
emulation as means to explore different task placement strategies.

\section{Workload and Runtime Emulation}
\label{sec:workload_emulator}

Distributed infrastructures encompassing the edge and cloud are complex,
heterogeneous, and dynamic, making top-down designs and resource allocation
decisions complicated. Thus, it is essential to evaluate potential decisions and
their impact on the performance before real-world deployment. To address this
issue, we use \emph{\radicaldreamer (Dynamic Runtime and Execution Adaptive
Middleware EmulatoR)}~\cite{radical_dreamer} to evaluate deployment modalities,
performance trade-offs and workload placement decisions for edge-to-cloud
applications~\cite{wlms_emulator}. In the following, we describe the fundamental
components and their interactions of the framework.

\begin{table}[t]
  \footnotesize
	\centering
	\begin{tabular}{|p{1.cm}|p{7.13cm}|}
	 \hline
	 
	 Task & Abstraction of a computational process to be executed. A task is characterized by  a \texttt{num\_ops}, i.\,e., the number of operations required to compute. Variability in the performance can be modeled using a statistical distribution and the \texttt{ops\_dist} parameter.\\ \hline

	 Workload & Collection of tasks submitted to the emulator at a time. \\ \hline %
	 
	 Resource & Description of the entity that is executing a given workload. A resource is defined by three parameters: the number of cores, each core's capacity defined as the number of operations per core, and a delay. Variability in the performance can be modeled using a statistical distribution. \\  \hline %

	 Resource Manager &Orchestrates emulation runs across multiple resources.  \\  \hline %
	 
	 Session & Accepts user-specified resource, workload, and scheduling configurations and manages emulation run via Resource Manager.\\  \hline

	\end{tabular}
	\caption{\textbf{\radicaldreamer Concepts:} The framework uses the concepts of \texttt{Task} and \texttt{Workload} to model the characteristics of an application (e.\,g., ensembles of heterogeneous/homogeneous tasks) and \texttt{Resource} to model distributed infrastructure. The \texttt{Session} and \texttt{ResourceManager} are used for the emulation execution.\label{tab:dreamerDescp}\up\up}
\end{table}

Table~\ref{tab:dreamerDescp} provides an overview of the concept used by the emulator. The emulator requires two inputs: (i) the workload and (ii) the resources. A workload is defined as a collection of tasks that are submitted to the emulator.  A resource's capacity is defined by the number of cores and the supported throughput on operations per second. Multiple resources can be grouped. Further, the emulator allows the modeling of delays, e.\,g., induced by network latencies.

The emulator is capable of accommodating both application and infrastructure variability. For example, an ensemble of heterogeneous tasks can be modeled using the \texttt{ops\_dist} attribute of a workload, which accepts different distributions, e.\,g., normal or uniform distribution. This feature is important to model, e.\,g., variabilities that arise in the processing times due to imbalanced datasets. Similarly, performance variability and heterogeneous resources can be emulated using the \texttt{perf\_dist} attribute of a resource.

The \texttt{Session} object functions as an entry point and accepts a workload, resource, and schedule configuration as input. The \texttt{ResourceManager} handles the emulation of the workload on the defined resources.   Session and Resource Managers communicate via a RabbitMQ service.   The \texttt{Session} records the execution and documents the results, e.\,g., the execution times of each task and the time-to-completion (TTC), so that these can be analyzed posterior to the experiment.

The emulator provides the ability to investigate different resource configurations, scheduling policies, application algorithms, parallelization, and distribution strategies. The costs per emulation run are very low, allowing broad explorations before setting up real-world systems and experiments.

\section{Experiments}
\label{sec:experiments}

\label{subsec:emulation_eval}

Insights into the complex and dynamically varying edge-to-cloud systems' performance characteristics are crucial to ensure appropriate response times, throughput and to optimize resource usage. In this section, we evaluate our emulation approach using a  machine learning
application using the K-Means algorithm. For this purpose, we compare the
performance of a real-world execution with the emulated performance.  We use the
XSEDE Jetstream cloud~\cite{jetstream} and two different virtual machines (VM)
types:  a single-core VM with 2\,GB of memory (edge), which is comparable to a
Raspberry Pie, and a 44 core VM with 120\,GB of memory (cloud).

The K-Means~\cite{kmeans} algorithm is a popular unsupervised machine learning algorithm that groups related data points in a dataset to discover underlying patterns. We model a K-Means application with the emulator and compare it to a real-world execution. For this purpose, we use the Scikit-Learn K-Means implementation with one core~\cite{scikit-learn}.
The complexity for a K-Means algorithm is $O(K N T)$, where $K$ is the number of clusters, $N$ is the number of points in the dataset, and $T$ is the number of iterations for a single run.

The K-Means application is translated to a workload of T tasks, where each task represents an iteration of the algorithm. Based on the number of points and clusters ($K N$) in each iteration, we compute the number of operations per task. Each iteration is then mapped to a task. The resource's capacity defined by the supported operations per second is derived from an XSEDE Jetstream cloud micro experiment. We then calculate the operation per second executed for the given input data size, i.\,e., from $32$ to $10^{6}$ points, and configure the emulator's workload and tasks accordingly.

\begin{figure}
	\centering
	\includegraphics[width=0.49\textwidth]{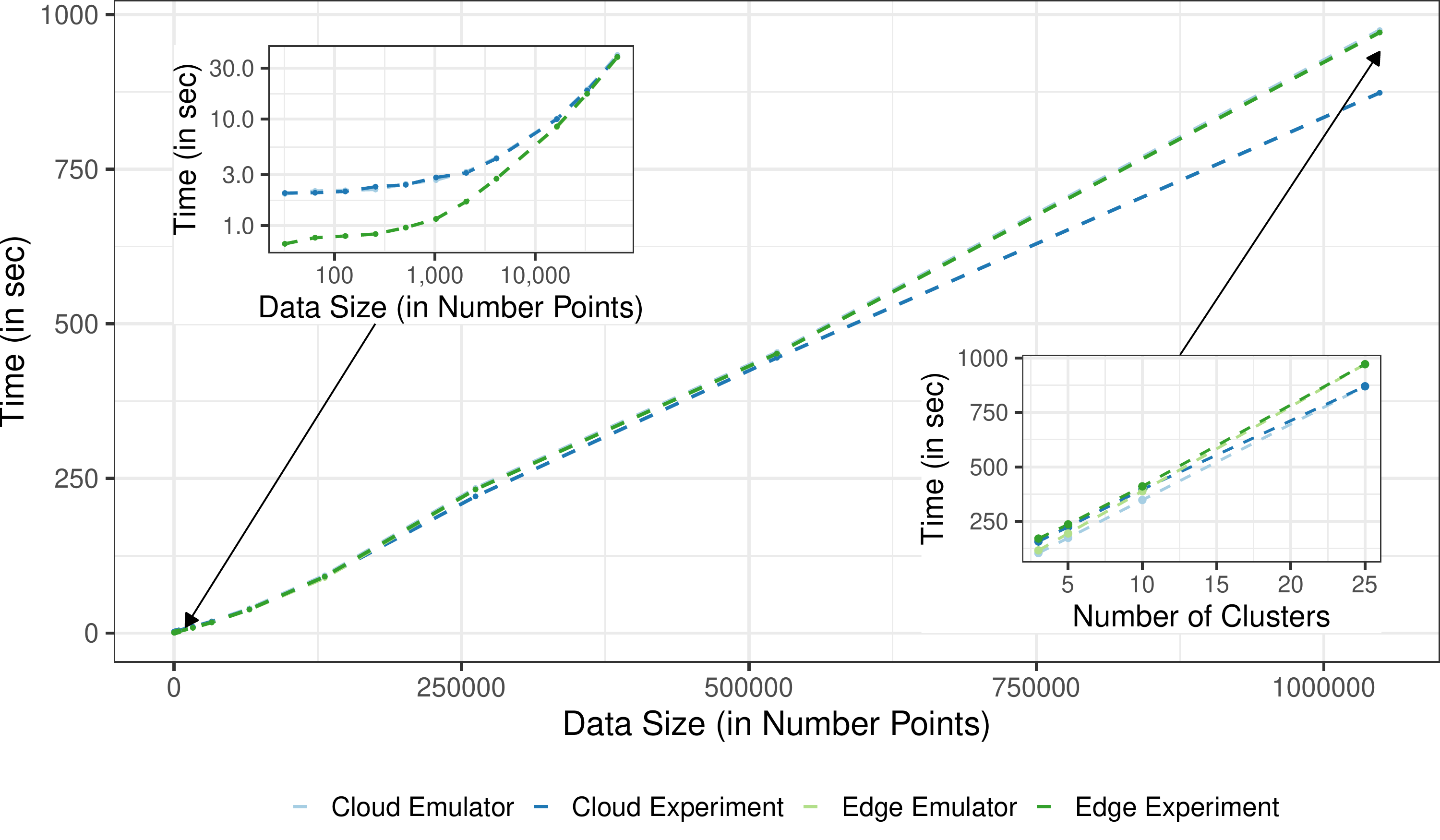}
	\caption{\textbf{Edge-to-Cloud K-Means -- Experiment and Emulation for Cloud- and Edge-Centric Deployment Modalities:} The main graph shows that the emulator captures the characteristics of K-Means well. The upper inset shows the behavior for smaller message sizes, indicating a small advantage for the edge. The lower inset shows TTC for different cluster numbers and $10^{6}$ data points. 
	\label{fig:kmeans_emulator_experiment}\up\up}
\end{figure}

We conduct an end-to-end experiment to evaluate  task placement strategies for an edge-centric and cloud-centric scenario. Further, we assess the quality of the emulated time-to-completion (TTC) by comparing it to real experiments. For the cloud scenario, we use a Kafka broker to move data from the edge to the cloud. 

Figure~\ref{fig:kmeans_emulator_experiment} shows the results of the evaluation. As shown, we can capture real-world behavior with our emulation approach. As seen in the upper inset, the limited edge resources can outperform the cloud for smaller data sizes ($<$30k points). In this case, the overhead for setting up data transfers to the cloud too large, particularly compared to the compute time. As the data size increases, the computational demands grow, making it advantageous to move the data to the cloud.

Further, we investigate the relationship between the number of clusters for K-Means and TTC using a configuration with a fixed $10^{6}$ data points (see lower inset in Figure~\ref{fig:kmeans_emulator_experiment}). As the cluster size increases, K-Means algorithm's complexity grows, leading to a longer TTC. In summary, both experiments validate that the used emulation approach is suitable for capturing important characteristics of real-world applications.

When comparing edge- vs. cloud-centric task placement strategies, the experiments
demonstrate that the right task placement has a significant impact on the overall performance. 
For smaller data sizes, we observed up to 65\,\% better performance  on the edge devices; for larger data sizes, the cloud runtime was up to 10\,\% faster.

\section{Related and Previous Work}
\label{sec:related}

Varshney/Simmhan~\cite{Varshney2019} provide an extensive survey of application
programming and scheduling models for the edge-to-cloud continuum.
Ref.~\cite{Ashouri_2019,6844677} review various emulation approaches for
specific aspects of the edge-to-cloud continuum that have emerged, e.\,g.,
focusing on network, compute, scheduling, or data. We focus on a narrow approach of using emulation for application-level task placement strategies.

CloudSim~\cite{cloudsim} is a well-known simulator for cloud infrastructures
and supports, e.\,g.,  federated infrastructures. The objective of CloudSim is
to support resource management and scheduling decisions. Several
extensions for edge computing emerged, e.\,g.,
IoTSim-Edge~\cite{jha2019iotsimedge} provides a comprehensive framework for
modeling edge-to-cloud continuum environments. It allows the
modeling of environments considering, e.\,g., (i) different IoT communication
protocols, (ii) edge device heterogeneity, (iii) data movements, and (iv)
energy.

While these emulation approaches are often feature-rich and support broad
objectives, they are typically focused on infrastructure, considering
application characteristics only on very abstract levels.  We specifically
address IoT and data-driven applications that require complex, distributed data
processing pipelines with highly dynamic data movements and computational
requirements. 

\section{Conclusion and Future Work}

Designing distributed applications that need to handle data and computing across multiple 
infrastructure layers is challenging. This paper presented a comprehensive
emulation approach for characterizing and understanding edge-to-cloud applications. 
Emulation allows exploring vast parameter spaces and
is an essential tool to guide experimental design, and experimentation is
crucial to evaluate end-to-end performance data across all layers. Additionally, the emulator 
allows the quick investigation of `what-if' scenarios. 

We demonstrated these capabilities by studying important resource management
challenges, e.\,g.,  the suitability of a machine learning model for a given
scenario. For example, in our experiments, we investigated the impact of model
complexity on latencies.

In the future, we will develop abstractions for distributed
edge-to-cloud applications that provide the means to devise and implement
effective resource management strategies. The envisioned Pilot-Edge abstraction is
based on the \pilotabstraction~\cite{pstar12}. It provides an easy-to-use
serverless Function-as-a-Service API that simplifies application development,
allowing developers to focus on application logic and application-level resource
management. It supports common patterns, e.\,g., cloud-centric, edge-centric,
and hybrid applications. Tasks can easily be moved to different parts of the
continuum, e.\,g., from the edge to the cloud, or vice versa.  Further, we will
tightly integrate our emulation approach with the Pilot-Edge middleware
providing the ability (i) to include parameters from real-world experiments and
(ii) to use the emulation to optimize the experiment's design.

\vspace{2mm}
\subsubsection*{Acknowledgments} 

\footnotesize We acknowledge Mikhail Titov for his work on \radicaldreamer.
Computational resources were provided by NSF XRAC award TG-MCB090174 and the
Leibniz Supercomputing Center (LRZ). This work was supported by a DOE SBIR
Award DE-SC0017047 to Optimal Solutions Inc.

\bibliographystyle{unsrt}
\bibliography{pilotedge,radical_publications,streaming}

\end{document}